# BEOL Electro-Biological Interface for 1024-Channel TFT Neurostimulator with Cultured DRG Neurons

Haobin Zhou[1], Bowen Liu[1], Taoming Guo[1], Hanbin Ma[2,3] and Chen Jiang[1*]

[1]Department of Electronic Engineering, Tsinghua University, Beijing, China, [2]CAS Key Laboratory of Bio Medical Diagnostics, Suzhou Institute of Biomedical Engineering and Technology, Suzhou, China, [3]Guangdong ACXEL Micro Nano Tech Company Ltd., Foshan, China.

*E-mail: chenjiang@tsinghua.edu.cn

**Abstract**

The demand for high-quality neurostimulation, driven by the development of brain-computer interfaces, has outpaced the capabilities of passive microelectrode-arrays, which are limited by channel-count and biocompatibility. This work proposes a back-end-of-line (BEOL) process for 1024-channel stimulator with bioelectrodes and waterproof encapsulation to stimulate dorsal root ganglion neurons. We introduce an active-matrix neurostimulator based on n-type low-temperature poly-silicon thin-film transistor, adding PEDOT:PSS and SU-8 as bioelectrodes and encapsulation. This enables precise stimulation of DRG neurons, addressing key challenges in neurostimulation systems.

## Introduction

Due to the advances in brain-computer interfaces [1], neurostimulation therapy [2], and related fields, the interaction between electronics to biological tissues is becoming more important. However, the conventional CMOS silicon-based electronics is highly integrated and rigid, which is difficult to achieve a flexible system for biological tissues. Thin film transistors (TFTs) have the advantage of mechanical flexibility and can be designed with high-channel-count neurostimulators to adhere to biological tissue over a large area, such as brain cortex. The circuit design of TFTs with four transistors and one capacitor (4T1C) enables high-channel-count active-matrix arrays for independent programming and simultaneous stimulation of 1024 channels [3, 4]. Additionally, this fabricated chip exhibits high yield, and it has the potential to achieve a fully flexible stimulation array. However, the chips fabricated from inorganic materials can be severely degraded in solution environments, and the interfaces of the chips are not suitable for biological tissues.

To address these issues, we introduce a BEOL process to fabricate a bioelectrode layer suitable for cell attachment along with a waterproof encapsulation layer to protect chip from degradation. A bioelectrode layer formed by the poly (3,4-ethylenedioxythiophene):polystyrene sulfonate (PEDOT:PSS) is spin-coated and photolithographically patterned on the chip to enhance the biocompatibility of the electro-biological interface [5, 6]. In addition, SU-8, a polymer material commonly employed for bioencapsulation, which possesses high biocompatibility, favorable mechanical properties, and superior waterproof performance, is used as a waterproof encapsulation layer [7]. Combined with the high-channel-count active array, this system enables neurostimulation for 100 μm resolution of dorsal root ganglion (DRG) neurons.

## LTPS Fabrication and Proposed Pixel Circuit

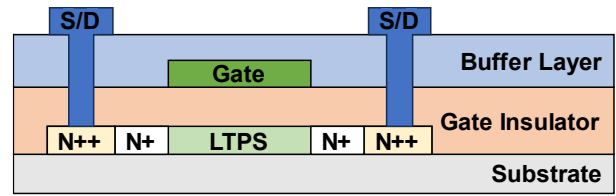

Fig. 1: Cross-sectional view of the fabricated n-type LTPS TFTs.

The proposed n-type low-temperature polycrystalline silicon thin-film transistors (LTPS TFTs) employ a top-gate structure, as illustrated in Fig. 1. The fabrication process started with the deposition of a poly-Si layer on a glass substrate. Following this, a channel-doping process was performed on the channel region, while the source and drain regions underwent an n-type doping process. Then, Mo was deposited to form the gate electrode, which was then patterned using standard photolithographic techniques. An $SiO_x$ buffer layer was deposited, followed by the etching of vias to establish connections to the source and drain metal Finally, a trilayer of Ti/Al/Ti was deposited to serve as the metal contacts for the source and drain electrodes.

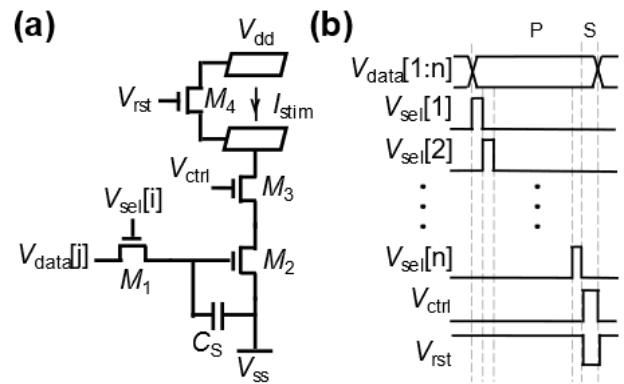

Fig. 2: (a) Schematic and (b) timing diagram of the proposed 4T1C stimulation pixel circuit. P and S stand for programming phase and stimulating phase in one cycle.

The neurostimulation pixel circuit includes 4T1C as shown in Fig. 2(b), in the programming phase, $V_{sel}$ changes to high voltage in sequence to turn on $M_1$ and $V_{data}$ changes to corresponding voltage. $M_1$ acts as a switch to program the voltages on data lines to the capacitor $C_s$ when the $V_{sel}$ is pulled up, and $C_s$ stores $V_{data}$ until the next programming phase. In the stimulating phase, $M_2$ transistor acts as a driving transistor to supply stimulation current to the electrodes according to the voltage stored in $C_s$. Both $M_3$ and $M_4$ transistors act as switches to turn on the corresponding branch circuit to perform the stimulation and discharge, respectively, when $V_{ctrl}$ and $V_{rst}$ are pulled up.

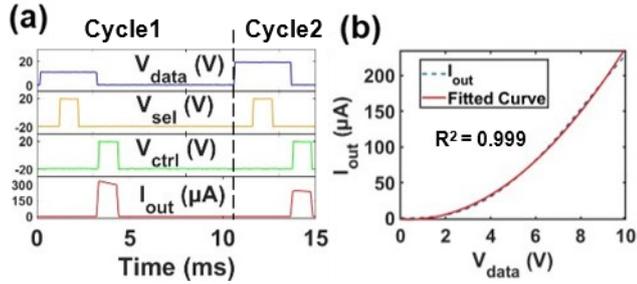

Fig. 3: (a) Measured output current $I_{out}$ and control voltage $V_{data}$, $V_{sel}$ and $V_{ctrl}$ of the pixel circuit for neurostimulation function and (b) Measured $I_{out}$ under different voltage of $V_{data}$.

We verified the neurostimulation function of the fabricated circuit, as shown in Fig. 3(a). It illustrates the outputs of the circuit, in which $V_{dd}$ was set at 20 V, $V_{sel}$ and $V_{ctrl}$ toggled between –20 V and 20 V. The capacitor $C_s$ was programmed with two different $V_{data}$ of 10 V and 19 V, respectively. Two output currents of 315 μA and 245 μA were observed, which are sufficient to evoke action potentials. $I_{out}$ is controlled by $V_{ctrl}$ and the amplitude can be modulated by $V_{data}$ effectively. Further, the $I_{out}$-$V_{data}$ relation is obtained by sweeping $V_{data}$ from 0 to 10 V, as shown in Fig. 3(b). The relation is quadratic, which corresponds to the voltage-controlled current relationship of a transistor in the saturation region.

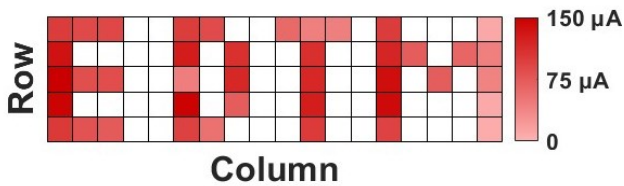

Fig. 4: The stimulation currents of 90 electrodes in the test region.

To apply patterned stimulation to DRG neurons, an area with 90 electrodes on the chip were selected, consisting of 41 electrodes programmed as stimulation pixels and the other 49 electrodes as non-stimulation pixels to form a pattern of 'EDTM'. The results of the outputs are depicted in Fig. 4, where the color intensity of each square represents the magnitude of the stimulation current. All the 41 stimulation electrodes successfully delivered stimulation currents, while all the non-stimulation electrodes did not exhibit current pulses, achieving a 100% yield. These results demonstrate the capability for precise stimulation at the targeted area.

### BEOL Process for Electro-Biological Interface

The BEOL process includes a PEDOT:PSS layer at the electrode sites and a waterproof encapsulation, as shown in Fig. 5(c). The PEDOT:PSS layer transformed the indium tin oxide (ITO) electrodes into bio-electrodes to improve biocompatibility [8]. The SU-8 encapsulation layer protects the electrical circuits and further enhance the biocompatibility of the surface. The bio-electrodes and encapsulation allow neurons to adhere and grow on the surface without additional coating process, as shown in Fig. 5(a).

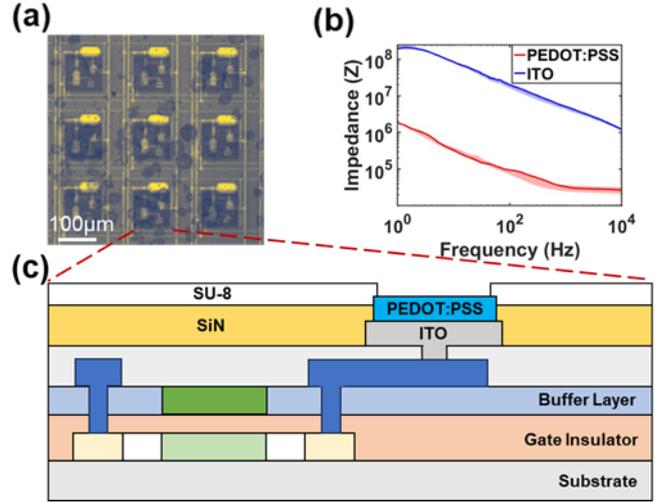

Fig. 5: (a) Cell adhesion and growth on the electro-biological interface. (b) Electrochemical impedance spectroscopy for ITO and PEDOT:PSS electrodes. (c) Cross-sectional view of the fabricated bio-electrodes and encapsulation.

Consequently, this process significantly reduced the contact resistance at the electrode interface. The electrodes are square-shaped with a side length of 100 μm. Electrochemical impedance spectroscopy of the bio-electrodes and the ITO electrodes (i.e., with and without PEDOT:PSS layer) is illustrated in Fig. 5(b). The impedance of the electrodes demonstrates typical frequency-dependent characteristics, i.e., decreasing with the increase of frequency. It is notable that the impedance of PEDOT:PSS/ITO electrodes are 1 to 2 orders in magnitude lower than that of ITO electrodes.

### Precise Neurostimulation of DRG Neurons

The DRG neurons were isolated from the spine of mice and cultured on a neurostimulation chip for 12 hours. Cells were stained with Fluo-4 AM calcium ion indicators during adherent growth and patterned programmed electrical stimulation [9]. It was possible to observe fluorescent images of calcium ions with a fluorescence imaging microscope, and thus the firing process of the neurons, as Fig.6 shown.

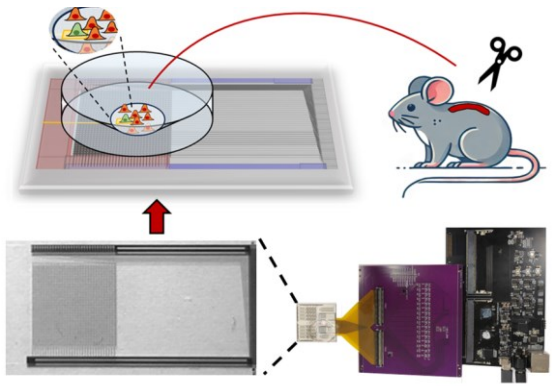

Fig. 6: Conceptual diagram of a system for neurostimulation of DRG neurons.

As Fig. 7(a) illustrated, the left two columns are stimulation electrodes, with stimulation applied from 5 to 30 seconds. In the stimulated region, the fluorescence images of neurons adhered to the electrodes exhibited a marked increase in brightness of calcium ion indicators during stimulation, followed by a decrease in intensity after the stimulation ended.

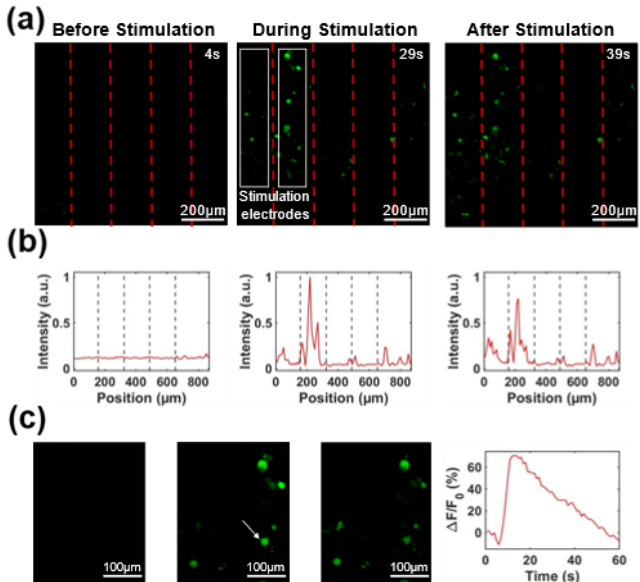

Fig. 7: Patterned electrical stimulation-induced action potentials observed through fluorescent calcium imaging. (a) Fluorescence images of neurons in regions which are programmed for current stimulation and no-current stimulation, with stimulation from the 5s to 30s. The dashed lines indicate electrode gap regions; The left two columns are stimulation electrodes, while the right side is non-stimulation region. (b) Lateral fluorescence intensity distribution across the fluorescence image. The dashed lines indicate electrode gap regions, aligned with Fig. 8(a). (c) Fluorescent images of a typical DRG neuron with 25s stimulation and calcium signal traces ($\Delta F/F_0$) over time.

Furthermore, the lateral fluorescence intensity distribution across the area in Fig. 7(a) was calculated and the results are shown in Fig. 7(b), indicating a significant increase in fluorescence intensity within the programmed stimulation region during the stimulation period, with values notably higher than those before and after stimulation. We selected a representative DRG neuron within the stimulated region and displayed its calcium signal traces ($\Delta F/F_0$) over time, as shown in Fig. 7(c).

## Conclusion

In conclusion, we performed BEOL on the high-channel-count neurostimulation array, adding a PEDOT:PSS layer to construct bio-electrodes and SU-8 as the waterproof encapsulation. This addresses the pressing needs in neurostimulation for biocompatibility and reliable signal transmission, providing high-channel-count and precise stimulation to neurons. We verified that the neurostimulation system enables patterned stimulation of DRG neurons and it holds potential for high-information-transfer brain-computer interface and electrical stimulation training of neural tissue. It may also be integrated with sensor arrays to form a closed-loop sensing and stimulation system.

## Acknowledgments

This work was supported in part by the National Natural Science Foundation of China (82151305 and 62374102). The authors gratefully acknowledge the assistance of TIANMA Microelectronics Corp. for manufacturing the chips.